\documentclass[preprint,12pt]{elsarticle}
\usepackage{graphicx}
\usepackage{bm}

\journal{Nuclear Instrument and Method A}

\begin{document}
\begin{frontmatter}
\title{Deuteron-induced reactions generated by intense Lasers \\for PET isotope production }
\author{Sachie Kimura$^a$}
\author{Aldo Bonasera$^{a,b}$}
\address{$^a$INFN-LNS, via Santa Sofia, 62, 95123 Catania, Italy}
\address{$^b$Cyclotron Institute, Texas A\&M University, College Station TX 77843-3366, USA}
\begin{abstract} 
We investigate the feasibility of using laser accelerated protons/deuterons for
positron emission tomography~(PET) isotope production by means of 
the nuclear reactions $^{11}$B($p,n$)$^{11}$C and $^{10}$B($d,n$)$^{11}$C.
The second reaction has a positive Q-value and no energy threshold. One can, therefore, make use of the lower energy part of 
the laser-generated deuterons, which includes the majority of the accelerated deuterons.    
The $^{11}$C produced from the reaction $^{10}$B($d,n$)$^{11}$C is estimated to be 7.4 $\times$ 10$^{9}$ per laser-shot at the Titan laser at Lawrence Livermore National Laboratory. Meanwhile a high-repetition table top laser irradiation is estimated to generate 3.5 $\times$ 10$^7$ $^{11}$C per shot from the same reaction. 
In terms of the $^{11}$C activity, it is about 2~$\times$~10$^4$~Bq per shot. 
If this laser delivers kHz, the activity is integrated to 1~GBq after 3 minutes. 
The number is sufficient  for the practical application in medical imaging for PET.


\end{abstract}

\end{frontmatter}

\section{Introduction}

Laser accelerated proton beams have been applied to produce positron emitting isotopes by several groups~\cite{spencer2, fritzler,ledingham,robson}.
Those groups investigate the feasibility of $^{11}$C isotope production rate by the laser-induced reaction $^{11}$B($p,n$)$^{11}$C. 
The reaction has a negative Q-value of -2.765~MeV and, therefore, requires large scale lasers like the VULCAN laser facility at the Rutherford Appleton Laboratory~(RAL).
Such large scale lasers are single shot.    
At current technology table-top lasers, which generate low temperature plasma but operate at high repetition rate, 
cannot lead to the reaction. In Ref.~\cite{spencer2} from the reaction $^{11}$B($p,n$)$^{11}$C induced by the laser-generated protons 
and the secondary boron target,
the activities of 2~$\times$~10$^5$~Bq of the isotope $^{11}$C have been measured.
They have carried out the experiments using the VULCAN laser, which delivers pulses 
of energies up to 120~J, duration 0.9-1.2~ps and a wavelength of 1053~nm. 
This laser pulse focused on a thin Al foil   
generates protons of energies as high as 37~MeV, from the contaminants on the foil target.
Lately, the experiment is repeated at the same VULCAN laser
at a pulse energy of 300~J with a pulse duration of 750~fs~\cite{ledingham},
a $^{11}$C activity of 6 $\times$~10$^6$~Bq has been reported.
On the other hand the activity required for PET is at least 200~MBq and ideally it is 800~MBq~\cite{spencer2}.

In this paper we investigate the feasibility of this application but using the reaction $^{10}$B($d,n$)$^{11}$C, instead. 
This reaction has positive Q-value. There is, therefore, no energy-threshold for the reaction~\cite{volkovitsky}. 
This means that smaller size lasers with higher-repetition rates can be used in contrast to the large-scale lasers,
though such lasers generate proton beams at lower energies.   
Laser generated deuterons from deuterated plastic layer on an Al foil target has been investigated only recently~\cite{petrov,davis,willingale,fuchs}.
A numerical investigation suggests that by eliminating surface contaminants, which contain protons, 
deuterons can be accelerated successfully, even though the laser pulse energy is relatively low. 
We compare estimates of $^{11}$C yields for the proton- and deuteron-induced reactions
irradiated by two types of lasers. First the Titan laser 
at the Jupiter Laser Facility at Lawrence Livermore National Laboratory~(LLNL)~\cite{hey}. 
The laser pulse delivered by this laser has energies up to 150~J, the pulse length of 0.6~ps and
the wave length 1054 nm~\cite{hey}. These are close to the parameters of the VULCAN laser
at the configuration in Ref.~\cite{spencer2}. The repetition rate of the Titan laser is 2-shots per hour. 
The other is a table-top laser, ASTRA Ti:sapphire laser facility~\cite{spencer} at RAL, 
with a repetition rate of 2~Hz,  a laser energy of 200~mJ, a pulse length of 60~fs and a wave length of 790 nm.

\section{Ion acceleration by lasers}
Several mechanisms are proposed to be responsible for the ion acceleration by interaction of high-intensity laser pulses
with thin foil targets.   The target normal sheath acceleration (TNSA) is one of those mechanisms and is thought to be 
the dominant mechanism at laser intensities 10$^{18}$ to 10$^{21}$ W cm$^{-2}$.
We assume the TNSA mechanism as the mechanism of the proton and deuteron acceleration. 
In the electric field of a laser, electrons on the front surface of the target quiver.
The oscillating motion of the electrons causes a force, which drives the electrons into the target normal direction, due to the oscillational magnetic field
perpendicular to the electrons' motion.  
The accelerated electrons penetrate through the target, 
ionizing the surface contaminants on the back, and form a dense sheath on the back. 
The charge separation between the escaping electrons and the ions left behind causes an electrostatic field 
which accelerates the protons in the hydrocarbon contaminant on the back surface.  
Under the TNSA mechanism, the protons and deuterons are expected to be accelerated by the same force $e\vec{E}$, where 
$e$ and $\vec{E}$ are the electron charge magnitude and the electric field due to the charge separation, respectively.  
The energy of the deuterons accelerated by the electrostatic field is determined by the charge of the deuteron which is the
same of protons.  Consequently the energy of the deuterons  
should be the same as the energy 
of the protons accelerated by the same field, while their velocities are different. 
On this account we expect that the deuteron energy spectra is similar to that of the protons. 

However, recently published experimental data~\cite{petrov,davis,willingale} show a controversial result. 
Experiments on the deuteron acceleration have been performed using T-cubed 
laser at the university of Michigan. This laser delivers  a pulse energy up to 6~J  in a 400~fs.  
The target was a 6~$\mu$m thick Al or 13~$\mu$m thick Mylar, coated with a layer of deuterated polystyrene 
either on the back or on the front or both side of the target. 
For the detection of accelerated ions, a Thomson parabola spectrometer has been used with a micro channel plate 
with a phosphor screen. 
The experimental observation that protons are accelerated preferentially to deuterons~\cite{petrov,willingale} is because  
the protons in the surface contaminants are on the top of the deuterated plastic layer.  
The protons, which are facing directly to the charge-separation electric field, see the maximum field strength, 
while the deuterons underneath the surface contaminant see the reduced electric field shielded by the protons.  
If the electrostatic field formed on the back surface is large enough,  
the layer of protons are ablated totally by the field. 
Then the ions of the target become exposed to the field
and are accelerated by the remaining electrostatic field
energy. 
 Another reason could be that if you have a mixture of different particles, the ones with lower mass leave the target first because of the higher velocity. 
Thus once they leave the target the electric field is reduced and the heavier ones see a lower field. 
Thus it is important to have pure targets to maximize the deuteron flux.  
The surface contaminants layer can be removed either by thermal heating~\cite{fuchs} or using a secondary laser 
to ablate the surface.
The use of high-repetition lasers has an advantage from this view point, since the contaminants of lower mass (protons) will be
removed in the repetitional pulses.  
 The removal of the contaminant allows acceleration of the target material~(deuterons) efficiently~\cite{davis,fuchs}.

The reductions both in the maximum deuteron energy and in the number of accelerated deuterons compared to the 
case of protons is observed to be approximately 1:10~\cite{willingale} at the given experimental condition, i.e.,
the thicknesses of the target and of the contaminant layer, and the specific laser parameters. 
Further experimental investigations are needed to understand difference between the proton- and deuteron-spectra, 
generated at different laser parameters and experimental conditions. 
For the purpose of the yield estimate of positron emitting nuclei, we use the same energy spectra as protons for deuterons 
at a set of given laser parameters.   
We stress that this alternative reaction to produce radioisotopes using deuterons is in any case interesting, considering 
the possibilities that with advancing technologies, higher repetition lasers of relatively low energy and short pulse might be feasible.

\section{Yield and angular distribution of neutrons}
\subsection{the reaction $^{11}$B($p,n$)$^{11}$C}

\begin{figure}
   \includegraphics[height=0.9\textheight, bb=0 0 612 792]{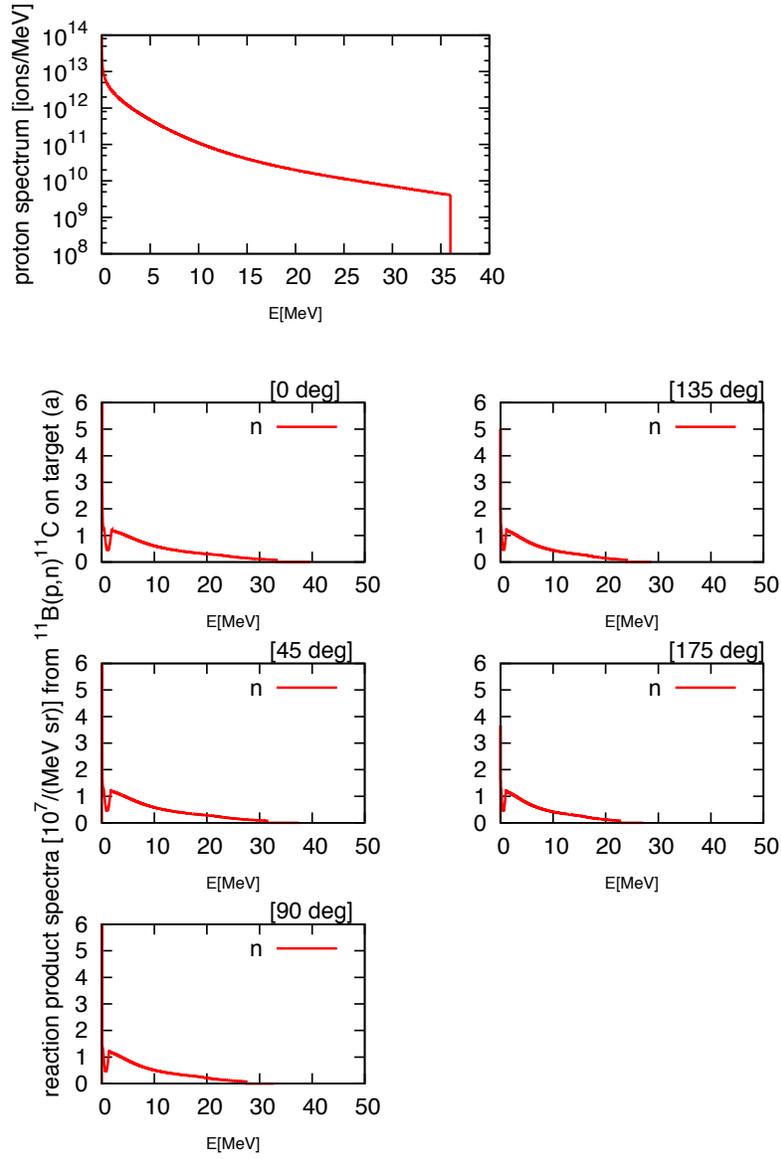}

   \caption{\label{fig:Pspec2} Proton spectrum at Titan laser irradiation on a Au target coated with ErH$_3$ (top panel).
   Expected neutron spectra for the reaction $^{11}$B($p,n$)$^{11}$C on a 3~mm-thick target at emission angles, 0, 45, 90, 135 and 175 degrees, at Titan laser irradiation.  }
\end{figure}

The yield from the reaction $^{11}$B($p,n$)$^{11}$C is estimated at Titan laser irradiation at LLNL.   
By irradiating Au foil target of thicknesses  from 5~$\mu$m to 250~$\mu$m, coated with 200 nm of ErH$_3$ on the rear surface, proton plasma 
which has the highest energy 36~MeV have been observed~\cite{hey}.
The proton spectra from this laser irradiation interacting with a 14~$\mu$m thick Au target coated with ErH$_3$ is measured with RC film. 
The spectra are parametrized by the following two temperatures distribution:
\begin{equation}
\label{eq:protonsp}
\phi(E)=N_1 e^{-E/T_1} \frac{1}{\sqrt{\pi T_1 E}}+N_2 e^{-E/T_2} \frac{1}{\sqrt{\pi T_2 E}},
\end{equation}
where $T_1=3.3$~MeV, $T_2=13.5$~MeV, $N_1=1.2 \times 10^{13}$ and $N_2=2.3 \times 10^{12}$~\cite{andy,hey}.
We reproduce the observed proton spectra in the top panel in Fig.~\ref{fig:Pspec2}.
If one locates a secondary boron slab target after the Au target, $^{11}$B in the target undergoes the nuclear reaction $^{11}$B($p,n$)$^{11}$C.
In this simulation, we assume that the secondary target is a 3 mm-thick slab of natural boron. The thickness of the target is chosen to be same as in the 
experiment at VULCAN laser~\cite{spencer2}. 
We call this target (a). Given that the proton spectra spread as high as 36~MeV,
the secondary target must be thick to ensure that all the accelerated protons which go through the 
boron target cause the nuclear reaction. 

The $^{11}$C yield~($N_f$) from the laser irradiation is evaluated by~\cite{volkovitsky} 
\begin{equation}
	\label{eq:nf}
	N_f=N_i \sigma(E) n_T d,
\end{equation}
where $N_i, \sigma(E), n_T$ and $d$ are the number of accelerated ions, the cross section of the reaction, the density
and the effective thickness of the boron target, respectively. 
The cross section for the reaction $^{11}$B($p,n$)$^{11}$C is shown in Appendix in Fig.~\ref{fig:cssf} by triangles and is roughly 100 mb at the energy 
higher than 5 MeV. 
The target density is $n_T$=8 $\times$ 10$^{22}$ cm$^{-3}$.
We replace the effective thickness of the target by the projected range of the protons in the target~\cite{srim}.
The projected range is given in a table as a function of the proton incident energy. 
For a rough estimate of the yield, 
we approximate the proton energy in the plasma, which spread from 0 to 36~MeV in reality, by the most probable energy~\cite{kb} for the 
reaction.  
The most probable energy for the plasma ions that cause the nuclear reaction is given by the Gamow peak energy 
($E_G$)~\cite{clayton} and
is found using the saddle point method, i.e.:
\begin{equation}
  \label{eq:gamow}
  \frac{d}{dE}\left(\frac{E}{kT}+bE^{-\frac{1}{2}} \right)=0,
\end{equation}
where $b=31.28Z_1Z_2 M^{\frac{1}{2}}$ (keV$^{\frac{1}{2}}$), denoting the atomic number of the colliding nuclei $Z_1, Z_2$ and 
the reduced mass number $M$. The reduced mass number $M=A_1A_2/(A_1+A_2)$, where $A_1$ and $A_2$ are the mass numbers of the colliding 
ions. 
The temperature $kT$ is replaced by the plasma temperature, then, 
\begin{equation}
  \label{eq:gamow2}
  E_G=\left(\frac{bkT}{2}\right)^{\frac{2}{3}}. 
\end{equation}
For the case of proton plasma generated by the Titan laser $E_G$ is 3.9~MeV.  At this energy the cross section of the reaction and 
the projected range of the protons are about 50~mb and $d=$150~$\mu$m, respectively.
A rough estimate of the fusion yield for the Titan laser facility gives 
\begin{equation}
	N_f=10^{13} 50 (mb)  8 \times 10^{22} (cm^{-3}) 150 (\mu m) \sim 5. \times 10^{8}, 
\end{equation}
per shot.
We note that the effective energy used in the estimate is higher than the threshold of the reaction.
  
For a more precise estimate of the yield,  
we calculate the angular distribution of the reaction products, i.e., neutrons, from the given proton spectra.    
From the kinematical relations the emission angle of $^{11}$C is limited to less than 60 deg. 
The energy of generated $^{11}$C is also limited to less than 3~MeV. 
Given that the emission angle is forward and the energy is very low, all $^{11}$C are stored in the target.
The thickness of the secondary target is enough to let $^{11}$C lose its energy inside the target.  
Experimentally the activity of $^{11}$C is measured by detecting 0.51~MeV gamma-rays by NaI scintillation detectors.    
For completeness we determine $^{11}$C yield from the neutron yield, which could bring other informations 
on the reaction mechanism in the plasma.

The angular distribution of the neutron yield is assessed by 
\begin{equation}
	\label{eq:dnf}
	\frac{\partial N_f}{\partial E \partial \Omega}= \phi(E) \frac{\partial \sigma(E)}{\partial E \partial \Omega} n_T d(E),
\end{equation}
where $\phi(E)$ and $\frac{\partial \sigma(E)}{\partial E \partial \Omega}$ are the proton energy spectra 
and the differential cross section of the reaction; the proton spectra is parametrized in Eq.(\ref{eq:protonsp}) and shown in the top panel in Fig.~\ref{fig:Pspec2};
$d(E)$ is the projected range which is a function of the incident 
energy, as mentioned before. 
We use the assumption of an isotropic angular distribution of neutrons in the center-of-mass (CM) system. 
The differential cross sections at different angles are obtained by transforming the cross sections from the CM system to the laboratory system.
In Fig.~\ref{fig:Pspec2} the lower panels show the neutron spectra at emission angles, 0, 45, 90, 135 and 175 degrees, 
from the proton energy distribution in the top panel. 
One can see that neutrons have broad energy spectra from 0 to 30~MeV, especially at the forward angle. 

By integrating the neutron yield, the total yield of the reaction $^{11}$B($p,n$)$^{11}$C is estimated to be 
1.5$\times$10$^9$ at Titan. This is in good agreement with the rough estimate given above.

\subsection{The reaction $^{10}$B($d,n$)$^{11}$C} 

$^{11}$C ions are generated by the deuteron-induced reaction $^{10}$B($d,n$)$^{11}$C, as well. 
This reaction has a positive Q-value 6.5~MeV.  
If one uses deuteron enriched hydrocarbon coating on a Au foil target and replace the secondary natural boron target by
a $^{10}$B enriched target, 
$^{10}$B undergoes the nuclear reaction  $^{10}$B($d,n$)$^{11}$C.
Experimental data of deuteron spectra at such a setup is not yet available. 
Assuming that the deuteron energy spectra is the same as the proton energy spectra, we estimate
the yield for $^{10}$B($d,n$)$^{11}$C similarly to the reaction $^{11}$B($p,n$)$^{11}$C. 
Eq.~(\ref{eq:nf}) gives again a rough estimate of the fusion yield by substituting $N_i$=10$^{13}$, $\sigma=200$ mb, $n_T=8 \times 10^{22}$ cm$^{-3}$ and $d=150~\mu$m, one obtains 
$N_f \sim 2 \times 10^{9}.$
 \begin{figure}
	 \includegraphics[height=0.7\textheight, bb=0 0 612 512]{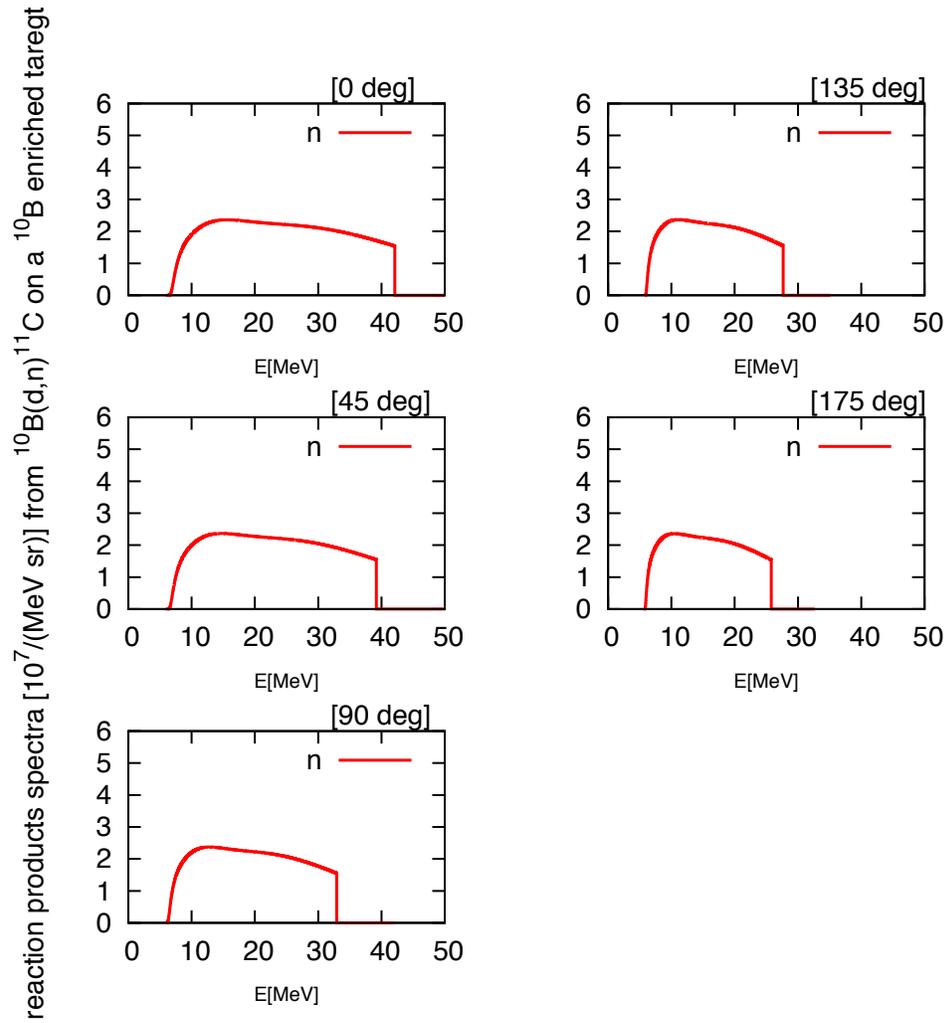}

   \caption{\label{fig:Pspec10Bdn} Expected neutron spectra from the deuteron-induced reaction $^{10}$B($d,n$)$^{11}$C 
   on the secondary 5~mm-thick $^{10}$B enriched target at 5 selected emission angles at Titan.}
\end{figure}
In Eq.~(\ref{eq:dnf}) by substituting the differential cross section by that of the reaction  $^{10}$B($d,n$)$^{11}$C, we estimate 
the neutron yield from the interaction between the deuteron plasma and the secondary $^{10}$B enriched target.  
Fig.~\ref{fig:Pspec10Bdn} shows the neutron yield estimated for the reaction $^{10}$B($d,n$)$^{11}$C at 5 selected emission angles. 
The neutron spectra range from 5.93~MeV, due to the reaction Q-value, to more than 20~MeV at all the detection angles.
We note that the secondary boron target is chosen to be 5~mm thick to ensure that all the deuterons which go through the 
boron target cause the nuclear reaction. 
By integrating the neutron yield,
the total yield of the reaction $^{10}$B($d,n$)$^{11}$C is 
estimated to be 
7.4 $\times$ 10$^9$ per shot at Titan. This is again in good agreement with the former 
rough estimate.
We also notice that neutrons could be produced by the deuteron disintegration. We expect this effect to be small because of the Q-value, but it could be precisely determined from kinematical considerations and from a precise measurement of the neutron yield and the gamma decay from $^{11}$C. 

\begin{figure}
   \includegraphics[height=0.9\textheight, bb=0 0 612 792]{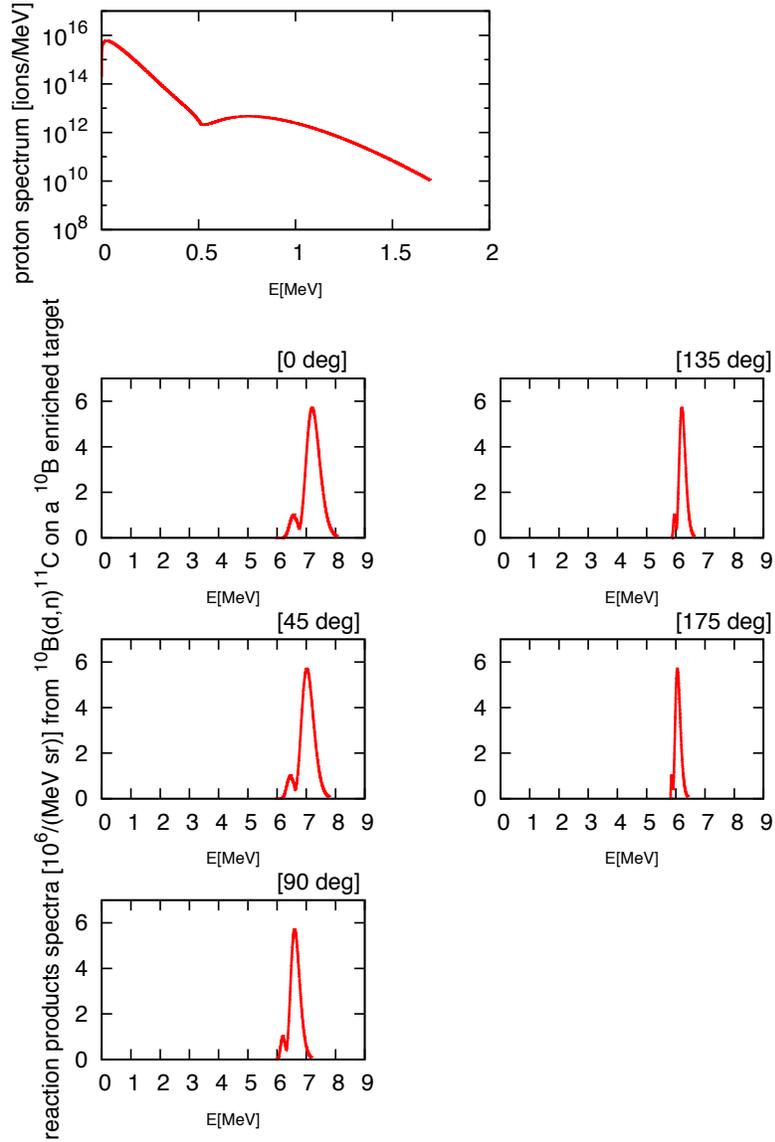}

   \caption{\label{fig:Pspec10Bdn62} Proton spectrum at a table-top laser irradiation on a 36~$\mu$m-thick Mylar target (top panel).
   Expected neutron~(red curves) and $^{11}$C~(blue curves) spectra from the reaction $^{10}$B($d,n$)$^{11}$C on 3~mm-thick $^{10}$B 
   enriched target at 5 selected emission angles at a high repetition-rate table-top laser irradiation. }
\end{figure}

\bigskip 
  Spencer et al.~\cite{spencer} studied proton spectra from plastic, aluminum and copper targets from a high-repetition-rate tabletop-laser irradiation.
The laser parameters in the experiment are the laser energy of 200~mJ, the pulse length of 60~fs and the wave length of 790 nm.    
It is of interest that the plastic target irradiation, at such laser parameters, generates higher energy protons and that the proton spectra 
have characteristic shape which cannot be approximated by a simple Maxwellian~\cite{kb}.  
 Thus we choose a proton spectra from Mylar target irradiation, which are shown in Fig. 2 in Spencer et al.~\cite{spencer}, 
 and fit the experimental data by the following formula:
  \begin{eqnarray}
  \label{eq:nn}
  \phi(E)&=&c\frac{1}{2\sqrt{\pi^3}}\frac{\sqrt{E}}{(kT_{HH})^{3/2}} \exp(-E/kT_{HH}) \nonumber \\
  &+&\frac{c_1}{4\pi}\frac{\sqrt{EE'}}{(kT_1)^{2}} \exp(-E'/kT_1), 
\end{eqnarray}
where $E$ is the energy of the protons in the laboratory system and
\begin{equation}
  E'= E-2\sqrt{EE_0}+E_0;
\end{equation}
$c, c_1,T_{HH}, T_{1}$ and $E_0$ are fitting parameters; $k$ is the Boltzmann constant.  
$E_0$ represents the energy of an extra acceleration as discussed in a previous work~\cite{kb}.
For the spectra of the irradiation on 36~$\mu$m-thick target, which has the number of accelerated ions larger than the other spectra, 
the obtained fitting parameters are listed in Tab.~\ref{tab:fit}.
The proton spectrum given by these parameters is reproduced in the top panel in Fig.~\ref{fig:Pspec10Bdn62}.   
 Assuming that the deuteron spectra in deuterated plastic target irradiation are similar to the proton spectra, 
we estimate the total yield for the reaction $^{10}$B($d,n$)$^{11}$C.
The angular distribution of the neutron yield with the secondary boron target is 
calculated by Eq.~(\ref{eq:dnf}).
The neutron spectra at 5 selected emission angles are estimated and shown in Fig.~\ref{fig:Pspec10Bdn62}.
The neutron spectra range from 5.9~MeV to 8.1~MeV at the forward angle and to 6.4~MeV at the backward angle. 
The evident two peaks in the neutron spectra are due to the two peaks in the proton spectra in the top panel. 
  \begin{table*}[htbp]
  \centering
  \begin{tabular}{c|ccccc}
    \hline
    T thick. & $c$ & $kT_{HH}$  & $c_1$ & $kT_1$ & $E_0$  \\
    ($\mu$m) & (ions/50 keV) & (keV) & (ions/50 keV)  & (keV) & (keV)  \\
    \hline
   36 & (4. $\pm$ 1.) 10$^{14}$ & 62.$\pm$3. & (6.7$\pm$2.) 10$^{10}$  & 41.$\pm$12. & 513.$\pm$62.  \\ 
    \hline
  \end{tabular}
  \caption{Fitting parameters for 36 $\mu$m Mylar target irradiation in Spencer et al~\cite{spencer}. }
  \label{tab:fit}
\end{table*}
By integrating the neutron yield,
the total yield of the reaction $^{10}$B($d,n$)$^{11}$C is estimated to be  
3.5 $\times$ 10$^7$ per shot.  
While in the reaction $^{11}$B($p,n$)$^{11}$C there will be no $^{11}$C yield at this table-top high-repetition laser irradiation.

\section{Discussion}

The total yield of the reaction $^{11}$B($p,n$)$^{11}$C is estimated to be 
1.5 $\times$ 10$^9$ per shot at Titan laser irradiation.
While the total yield of the reaction $^{10}$B($d,n$)$^{11}$C is estimated to be 
7.4 $\times$ 10$^9$ per shot at the same laser irradiation.
The estimated yield of the reaction $^{10}$B($d,n$)$^{11}$C is five times higher than that of the other reaction, given the same laser-pulse irradiation condition. 
 This yield gain by changing the reaction is explained by the fact that the deuteron-induced reaction has a positive Q-value, 
while the other proton-induced reaction has a negative Q-value. That is a large part of the plasma ions, which have energy less than 3 MeV, 
do not contribute to the yield in the reaction $^{11}$B($p,n$)$^{11}$C. 
If we use a high repetition-rate tabletop-laser, the yield is estimated to be
3.5 $\times$ 10$^7$ per shot from the reaction $^{10}$B($d,n$)$^{11}$C.
We note that the yield for the reaction $^{10}$B($d,n$)$^{11}$C could be overestimated due to 
our assumption of the identical energy spectra for protons and deuterons.  

If the major part of ions accelerated by this laser have energies less than 0.5~MeV,  
while the cross section of the reaction $^{10}$B($d,n$)$^{11}$C is small at the incident energy less than 0.5~MeV, as shown by diamonds 
in Fig.~\ref{fig:cssf}, 
the amount of $^{11}$C yield might be increased by using a more intense high-repetition laser. 
Here we assume a high-repetition laser, which meets the above requirement, 
at a laser energy 1~J and a pulse duration 40~fs and a repetition rate 10~Hz, and estimate the yield from this laser. 
These laser parameters are, indeed, similar to the parameters of the high repetition laser in Fritzler et al.~\cite{fritzler}.  
By studying laser-driven proton scaling laws~\cite{fuchs_nature,robson_nature,kb-sclny}, 
we estimate the plasma temperature of 10~MeV and the number of accelerated ions of 10$^{11}$ for this laser. 
Moreover we assume the maximum proton energy of 10~MeV.
For the proton spectra characterized by these parameters, the yield for the reaction $^{11}$B($p,n$)$^{11}$C
and for $^{10}$B($d,n$)$^{11}$C are estimated to be 1.3 $\times$ 10$^6$ and 2.8 $\times$ 10$^6$, respectively, per shot.

Spencer et al. ~\cite{spencer2} and Fritzler et al.~\cite{fritzler} discuss the application of the laser generated proton plasma to the PET isotope production. 
Spencer et al. ~\cite{spencer2} used the reaction $^{11}$B($p,n$)$^{11}$C at VULCAN laser, while Fritzler et al.~\cite{fritzler}
used a table top laser, at a laser energy  of 840~mJ and a pulse length of 40~fs, and at a repetition rate of 10~Hz. 
 Spencer et al. ~\cite{spencer2} observed $^{11}$C activities of 2$\times$10$^{5}$~Bq per shot and
estimates 10$^9$~Bq activities after 500~s(8.3 minutes) of irradiation, assuming 10~Hz operation of VULCAN laser.   
Fritzler et al.~\cite{fritzler} estimated the activity 13.4~MBq after 30 min irradiation at the laser at Laboratoire d'Optique Appliquee~(LOA).  
 
The result in Spencer et al. ~\cite{spencer2} could be compared with our estimate at Titan laser irradiation 
(1.5$\times$10$^9 \times$ 0.69/20.4/60=850~kBq) using the same reaction. 
The difference comes from the fact that proton energy spectra from Titan laser-irradiation~(Top panel in Fig.~\ref{fig:Pspec2}) 
has a higher flux than that from VULCAN laser-irradiation~(Fig.~4 in Ref.~\cite{spencer2}). 
The increase of the proton flux could be caused by the usage of the Au target coated with ErH$_3$ in the experiment at LLNL,
while protons originate only from impurities on the target in the experiment at VULCAN.   
In the case of the reaction $^{10}$B($d,n$)$^{11}$C at Titan laser irradiation the produced $^{11}$C activity is estimated to be (7.4$\times$10$^9 \times$ 0.69/20.4/60=4200~kBq). 
The estimated activity is still much smaller than the required activity for PET isotope, i.e. 800~MBq.

 If one uses the reaction $^{10}$B($d,n$)$^{11}$C and the table-top, high-repetition laser, 
we have estimated $^{11}$C production of 3.5 $\times$10$^{7}$ per shot, in terms of the $^{11}$C activity 
it is (3 $\times$10$^{7}$ $\times$ 0.69/20.4/60 $\sim$ 2 $\times$ 10$^4$~Bq) per shot. 
Considering that $^{11}$C decays with the half-life of 20.4 min and given 2~Hz repetition-rate, after 30 minutes of irradiation
the activity is integrated to be 22~MBq.  
If this laser delivers kHz, the activities are integrated to 1~GBq after 3 minutes 
and meet the required activity for the practical use of the PET isotope production.

Alternatively, for another high-repetition laser which delivers higher energy pulses of 1~J and a pulse duration of 40~fs
and a repetition rate of 10~Hz, the $^{11}$C activities for the reactions $^{11}$B($p,n$)$^{11}$C and $^{10}$B($d,n$)$^{11}$C 
are estimated to be (1.3 $\times$10$^{6}$ $\times$ 0.69/20.4/60 $\sim$ 730~Bq) and 
(2.8 $\times$10$^{6}$ $\times$ 0.69/20.4/60 $\sim$ 1.5~kBq), respectively, per shot.
Given 10~Hz repetition-rate, after 30 minutes of irradiation
the activities are integrated to be 8.2~MBq and 17~MBq, respectively, for the reactions 
$^{11}$B($p,n$)$^{11}$C and $^{10}$B($d,n$)$^{11}$C.

We remind that 
the $^{11}$C activities from the reaction $^{10}$B($d,n$)$^{11}$C at given laser parameters are estimated
under the assumption of identical proton- and deuteron-energy spectra. 
Taking into account the recent experimental observation~\cite{willingale}, the $^{11}$C activities from
the reaction $^{10}$B($d,n$)$^{11}$C are likely to be reduced than the above estimates.  

\section{Summary}

We have studied  the feasibility of $^{11}$C isotope production by means of the nuclear reactions $^{11}$B($p,n$)$^{11}$C and
$^{10}$B($d,n$)$^{11}$C induced by laser-accelerated protons/deuterons.  
We have estimated $^{11}$C produced by the reaction $^{11}$B($p,n$)$^{11}$C to be 1.5 $\times$ 10$^{9}$ per shot at Titan laser irradiation. 
While $^{11}$C produced by the reaction $^{10}$B($d,n$)$^{11}$C at the same laser irradiation is estimated to be 7.4 $\times$ 10$^{9}$.
The latter deuteron-induced reaction increases the yield 5 times than that of the proton-induced reaction. 
However the yield per shot is still too low for the practical use in the PET isotope production. 

The yield of $^{11}$C at a table top, high-repetition laser has been estimated to be 3.5 $\times$ 10$^{7}$ per shot for 
the deuteron-induced reaction. In terms of the $^{11}$C activity, it is about 2 $\times$ 10$^4$~Bq per shot. 
If this laser delivers kHz, the activity is integrated to 1~GBq after 3 minutes 
and meet the required activity for the practical use of the PET isotope production.  

The $^{11}$C activities for the reaction $^{10}$B($d,n$)$^{11}$C are likely to be overestimated due to the assumption 
of the identical energy spectra for protons and deuterons.   
The realization of this potential application requires further research to understand optimum parameters 
for deuteron acceleration by intense lasers. 
Our results have demonstrated a promising alternative way of using 
high-intensity laser for the application in the PET isotope production.  

\bigskip 

\noindent
Acknowledgment
\

The hospitality and the financial support of Texas A\&M University are gratefully acknowledged by one of the authors (S.K.).

\appendix
\section{Cross sections}
\begin{figure}
   \includegraphics[height=.6\textheight, bb=0 0 846 594]{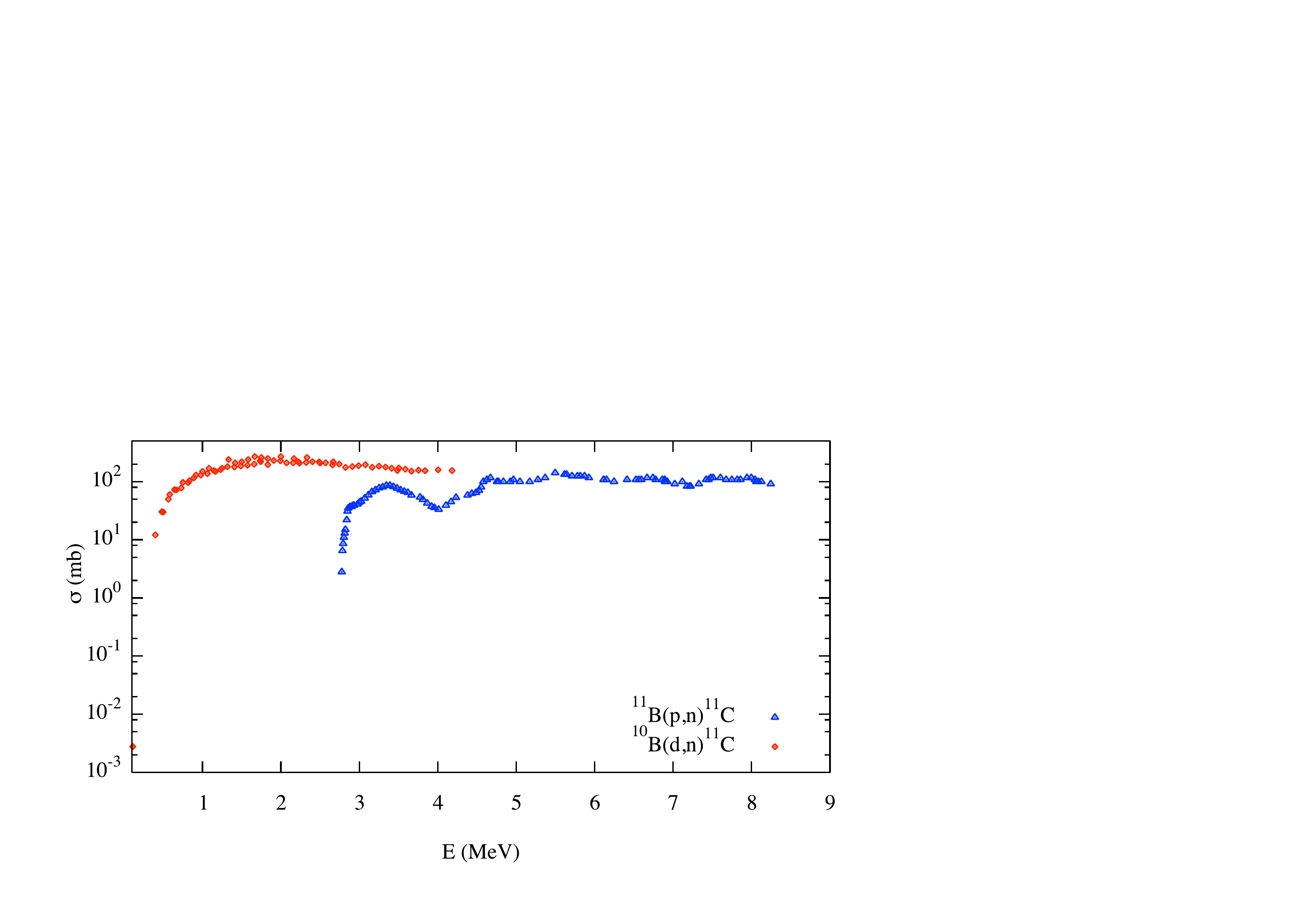}

   \caption{\label{fig:cssf} Experimental data of cross sections for the reactions $^{11}$B($p,n$)$^{11}$C~\cite{nacre}~(triangles) and $^{10}$B($d,n$)$^{11}$C~\cite{exfor}~(diamonds) are shown as functions of the incident energy both in the center-of-mass system. }



\end{figure}

Experimental data of cross sections for the reactions $^{11}$B($p,n$)$^{11}$C and $^{10}$B($d,n$)$^{11}$C are 
compared in Fig.~\ref{fig:cssf}. 

\bibliographystyle{model1-num-names}
\bibliography{pet}
%

\end{document}